\documentclass[12pt,preprint]{aastex}

\newcommand{\PG}{PG\,1211+143}

\newcommand{\asca}{{\it ASCA}}

\newcommand{\Msun}{\hbox{$\rm\thinspace M_{\odot}$}}
\newcommand{\ls}
{\mathrel{\hbox{\rlap{\hbox{\lower4pt\hbox{$\sim$}}}\hbox{$<$}}}}
\newcommand{\gs}
{\mathrel{\hbox{\rlap{\hbox{\lower4pt\hbox{$\sim$}}}\hbox{$>$}}}}
\def\Msun{\hbox{$\rm ~M_{\odot}$}}

\received{}
\begin{document}
\title{Evidence for Gravitational Infall of Matter onto the 
Super-Massive Black Hole in the quasar PG 1211+143?}
\shorttitle{Gravitational Infall of Matter in PG\,1211+143}
\shortauthors{Reeves et al.}
\author{J.N. Reeves\altaffilmark{1,2}, K. Pounds\altaffilmark{3}, 
P. Uttley\altaffilmark{1}, S. Kraemer\altaffilmark{1,4}, 
R. Mushotzky\altaffilmark{1}, T. Yaqoob\altaffilmark{1,2}, 
I.M George\altaffilmark{1,5}, T.J. Turner\altaffilmark{1,5}}

\altaffiltext{1}{Exploration of the Universe Division,
NASA Goddard Space Flight Center, Greenbelt Road, Greenbelt, MD 20771, USA; 
jnr@milkyway.gsfc.nasa.gov, pu@milkyway.gsfc.nasa.gov, 
richard@milkyway.gsfc.nasa.gov}
\altaffiltext{2}{Department of Physics and Astronomy, Johns Hopkins 
University, N Charles Street, Baltimore, Maryland, USA; 
yaqoob@skysrv.pha.jhu.edu}
\altaffiltext{3}{Department of Physics and Astronomy, University of 
Leicester, University Road, Leicester LE1 7RH, UK; kap@star.le.ac.uk}
\altaffiltext{4}{Department of Physics, Catholic University of America, 
620 Michigan Avenue NE, Washington, DC 20064, USA; 
stiskraemer@yancey.gsfc.nasa.gov}
\altaffiltext{5}{Joint Center for Astrophysics, University of Maryland 
Baltimore County, 1000 Hilltop Circle, Baltimore, MD 21250, USA; 
george@milkyway.gsfc.nasa.gov, turner@milkyway.gsfc.nasa.gov}
\begin{abstract}

We report the detection of redshifted iron K$\alpha$ 
absorption lines in the Chandra LETG spectrum of the 
narrow-line quasar, \PG. The absorption lines are observed at 4.22\,keV 
and 4.93\,keV in the quasar spectrum, corresponding to 4.56\,keV and 
5.33\,keV in the rest frame of \PG. From Monte Carlo simulations, the 
chance probability of both lines 
being false detections is low at $1.36\times10^{-4}$. 
Highly redshifted ionized iron K$\alpha$ (Fe~\textsc{xxv} 
or Fe~\textsc{xxvi}) 
is the most plausible identification for the lines at their observed energies. 
If identified with H-like iron K$\alpha$ at 6.97 keV, then the relativistic 
velocity shifts required are 0.40c and 0.26c. The extreme velocities 
can be explained by pure gravitational redshift if the 
matter exists in a stable orbit within 6 gravitational 
radii of the black hole. This would require a Kerr metric for the black hole. 
Alternatively the absorption may be the result of matter infalling directly 
onto the black hole, with a maximum observed velocity of 0.38c at 
$6R_{\rm g}$ in the Schwarzschild metric. 
This matter may originate in a failed outflow or 
jet, which does not escape the gravitational potential of the black hole. 

\end{abstract}


\section{Introduction}

One prime motivation of high energy research into Active Galactic 
Nuclei is to find evidence for the putative massive black hole, 
through the gravitational redshift imparted on photons emerging within a few 
gravitational radii of the event horizon. 
To date, most of the effort has focused on the 
iron K$\alpha$ emission line, predicted to show a relativistically
broadened profile \citep{Fabian89, Laor91}, which 
appeared to be common among Seyfert 1 galaxies, as observed by ASCA 
\citep{Tanaka95, Nandra97}. However the situation emerging from the  
newer XMM-Newton and Chandra data is more ambiguous, the spectra 
showing that narrower (non-relativistic) lines 
are commonplace \citep{Yaqoob04}. Indeed the red-wing 
of the broad iron line is often hard to distinguish from the 
effects of the warm absorber \citep{Reeves04, Turner05a}. 

Given the ambiguity in modeling the broad iron line, 
additional diagnostics of the inner disk material are needed.  
Narrow, transient and redshifted iron emission lines of 
modest statistical significance 
have recently been claimed in several AGN \citep{Turner02, Dovciak04, 
Porquet04} possibly originating from local hot-spots on the disk surface. 
Potentially  these lines can be tracked to reveal orbital motions in the 
inner disk \citep{Dovciak04, Turner05b}. However little evidence has 
been found to date for infall of matter onto the black hole. 
In this paper, we present evidence of highly redshifted iron K$\alpha$ 
absorption lines in the narrow-line Seyfert 1, \PG\ with Chandra LETG. 
This may either be interpreted as infall onto the massive black hole,  
or gravitational redshift occurring within $<6R_{\rm g}$ of a Kerr black 
hole.  \PG\ (at z=0.0809) is a prototypical `Big Blue Bump' quasar, 
and one of the brightest AGN in the soft X-ray band \citep{Walter93}. 
A previous XMM-Newton observation of \PG\ showed blue-shifted absorption 
from a high velocity ($v\sim0.1c$) outflow \citep{Pounds03}.  
The soft X-ray properties of
\PG, which confirm the blue-shifted absorber, will 
be presented in a subsequent paper. 
Values of H$_{\rm 0}$=70\,km\,s$^{-1}$\,Mpc$^{-1}$,
$q_{\rm 0}$=0.0 and $\Lambda_{\rm 0}=0.73$ are 
assumed and errors are quoted at 90\% 
confidence ($\Delta\chi^{2}=2.7$), for 1 parameter of interest. 

\section{Chandra and XMM-Newton Observations of PG 1211+143}

PG 1211+143 was observed by Chandra with the Low Energy Transmission Grating 
(LETG) in 3 consecutive orbits between 
21 June 2004 and 26 June 2004 of approximately 45\,ks exposure per orbit 
(see Table 1). An XMM-Newton observation was conducted simultaneously with the 
first Chandra orbit, but due to high and variable background, this was 
reduced to only $\sim$20\,ks useful exposure. 
The ACIS-S detector was used on the focal plane, with the S3 chip being 
at the telescope aimpoint. Chandra LETG 
source spectra, background spectra and response matrices were generated 
using CIAO 3.2.1. Streaks were removed from the ACIS-S CCD array, while 
events with grades 0, 2, 3, 4, and 6 were selected. After screening 
an exposure of 133.6 ks was obtained. 
Spectra and responses from the $\pm1$ orders were extracted and  
combined to maximize signal to noise. The time-averaged combined first order 
LETG count rate was 0.31 counts~s$^{-1}$. Data 
from the zeroth order image were not used, due to photon pile-up. The 1st order
spectra were binned so that one bin represents the FWHM resolution 
of the LETG ($\Delta\lambda=0.05$), ensuring sufficient counts per bin 
to use the $\chi^{2}$ minimization spectral fitting technique.

\section{LETG Spectral Analysis}

To parameterize the X-ray continuum, the LETG spectrum was fitted with 
a broken power-law model, modified by neutral Galactic absorption of 
$N_{\rm H}=2.8\times10^{20}$~cm$^{-2}$ \citep{Dickey90}, 
over the 0.4-8.0 keV bandpass.  
The soft and hard X-ray photon indices, break energy and model 
normalization were allowed to vary; the best fit values are 
$\Gamma_{\rm soft}=2.77\pm0.03$, $\Gamma_{\rm hard}=1.96\pm0.05$, 
break energy $E=1.7\pm0.1$~keV and normalized flux  
of $1.6\pm0.1\times10^{-3}$~photons~keV$^{-1}$~cm$^{-2}$~s$^{-1}$ 
(at 1 keV). 
The integrated flux is $1.42\times10^{-11}$~erg~cm$^{-2}$~s$^{-1}$ 
(0.4-8\,keV), corresponding to an unabsorbed X-ray luminosity of 
$2\times10^{44}$~erg~s$^{-1}$. 
The spectral fit obtained is formally 
unacceptable, with $\chi^{2}/{\rm dof}=686/544$ 
(dof=degrees of freedom) for a rejection probability 
of $2.9\times10^{-5}$. This is due to several ionized emission and 
absorption lines (e.g. from O, Ne) below 1 keV, 
although the continuum is fitted well above 2 keV. A subsequent paper 
will describe the soft X-ray spectrum of \PG\ obtained with the 
LETG (Reeves et al. 2005, in preparation). 

In this paper we concentrate on the spectrum in the iron K band above 2 keV. 
A single power-law model 
provides an adequate fit to the continuum above 2 keV, with a photon index 
of $\Gamma=1.99\pm0.06$. The fit statistic obtained 
is $\chi^{2}/{\rm dof}=116.4/93$, which is marginally acceptable 
(the rejection probability is $5\times10^{-2}$). 
Figure 1 shows the LETG counts spectrum above 2 keV, the lower 
panel showing the $\chi^{2}$ residuals of the data compared to the 
continuum model. Each bin 
represents the FWHM resolution of the LETG. Two strong negative deviations 
that contribute significantly towards the total $\chi^{2}$ 
are observed near 4.2 and 4.9 keV, due to two absorption 
lines in the \PG\ spectrum. To test this, two unresolved 
($\sigma=10$~eV) Gaussian absorption lines were added to the continuum model. 
A significant improvement 
in the spectral fit was obtained ($\chi^{2}/{\rm dof}=83.4/89$), 
with addition of 
two absorption lines at $4.22\pm0.03$~keV and $4.93\pm0.03$~keV (observed 
frame) with equivalent widths of $35\pm16$~eV and $57\pm23$~eV. 
The improvement in fit statistic upon adding each line was 
$\Delta\chi^{2}=13.8$ and $\Delta\chi^{2}=19.9$ respectively, while the  
overall fit statistic is now acceptable.  

To constrain the width of the two absorption lines, the LETG spectrum 
was rebinned to their maximum resolution, with each bin 
representing 1/4 FWHM resolution of the LETG ($\Delta\lambda=0.0125$\,\AA\ or 
$\Delta E=16$~eV at 4 keV). The rest-frame spectrum at this resolution 
near the absorption lines is shown in Figure 2. The 
spectral fit was minimized 
using the C-statistic \citep{Cash79}, due to the lower number of 
counts per bin at this resolution. The best fit width is $\sigma\sim21$\,eV
(at 4.9 keV), corresponding to a FWHM velocity of 
$\sim3000$\,km\,s$^{-1}$. However the line width is poorly constrained, 
the 90\% upper limit being $<7800$\,km\,s$^{-1}$. 
The variability of the absorption lines was tested by comparing the LETG 
spectra across all three Chandra orbits. The first LETG orbit was 
simultaneous 
with XMM-Newton, so the LETG and EPIC-pn spectra were fitted jointly. 
The results are shown in Table~1, while the C-statistic 
was used to minimize the spectral fits. 
The 4.2 and 4.9 keV lines are only formally 
detected during the 2nd orbit, although orbits 2 and 3 are statistically
consistent. The lines do appear to vary in equivalent width between 
orbits 1 and 2 at $>90\%$ confidence. 
 


\subsection{Statistical Significance of the Absorption Lines}

In order to check that the absorption is not due to calibration effects, 
the \PG\ spectrum was compared to all the archived LETG observations of 
3C\,273.  
No residuals are present in the summed or individual 3C\,273 spectra 
near the line 
energies 4.22\,keV and 4.93\,keV (Figure 3). The limit on the equivalent 
width of any absorption lines is $<4$~eV in the summed spectrum. 
The source and background spectral extraction regions of \PG, in each of the 
three Chandra orbits, were also inspected. No anomalous pixels or 
rows were observed in the CCD array and no background sources were present 
within the extraction regions. The separate $\pm1$ order spectra of \PG\ 
were also found to be consistent. Thus the lines are not due to systematic 
or calibration effects in the LETG. 


In order to assess the statistical significance of the lines, we 
wish to determine the chance probability of falsely detecting two 
lines, over an energy band of interest. In order to estimate this, 1000 
LETG spectra were simulated, under the null hypothesis assumption that 
the spectrum consists of a broken power-law continuum from 0.4-8.0\,keV, 
with no spectral lines.   
The distribution of $\Delta\chi^{2}$ values obtained 
by fitting a single Gaussian line to each of the 1000 spectra simulated in the 
null hypothesis case can then be compared to the measured 
$\Delta\chi^{2}$ in the real dataset to calculate the false 
probability of detection \citep{Porquet04}. 
The 1st order LETG spectra were simulated using identical 
continuum parameters and exposure as in the real \PG\ dataset
and binned at the FWHM resolution of the LETG.  
A single narrow Gaussian line (either in absorption or emission) 
was added to the continuum 
model and then fitted to each simulated spectrum. 
The line energy was stepped in increments of one FWHM resolution bin 
over the 3--7 keV band and the simulated spectrum was 
re-fitted at each increment to minimize $\chi^{2}$.  
The minimum $\chi^{2}$ obtained from adding the line to the spectral fit 
was compared to the $\chi^{2}$ obtained for the continuum model alone 
for each spectrum, in order to compute the distribution 
of $\Delta\chi^{2}$ versus probability for all 1000 simulated spectra. 
The 3--7 keV band was chosen for the line search as it is most likely
that a strong iron K$\alpha$ line will be observed over 
this energy range; i.e. $<3$\,keV is outside 
the iron K band-pass for any feasible redshift, while the LETG 
has little effective area above 7 keV. 
The improvement in fit statistic obtained for adding the two lines to the 
actual dataset was $\Delta\chi^{2}=13.8$ and $\Delta\chi^{2}=19.9$ 
for each line. The simulations show that the  
false probability of detecting a line with equal or greater
$\Delta\chi^{2}$ is 0.034 and 0.004 for the 4.2 and 4.9\,keV lines 
respectively. The probability of both lines being false detections is the 
product of these probabilities, which is $1.36\times10^{-4}$. 

As an independent check, we calculate the line significances using the 
actual errors on the fitted parameters. The absorption line fluxes 
at 4.2 and 4.9\,keV are 
$-(4.9\pm1.3)\times10^{-6}$\,photons\,cm$^{-2}$\,s$^{-1}$ and 
$-(5.9\pm1.5)\times10^{-6}$\,photons\,cm$^{-2}$\,s$^{-1}$ respectively, 
while the continuum fluxes at these energies are 
$1.40\pm0.10\times10^{-4}$\,photons\,cm$^{-2}$\,s$^{-1}$\,keV$^{-1}$
and $1.04\pm0.08\times10^{-4}$\,photons\,cm$^{-2}$\,s$^{-1}$\,keV$^{-1}$. 
The equivalent widths of the lines are then $-35\pm10$\,eV and 
$-57\pm15$\,eV respectively. The errors are $1\sigma$ values and the 
line flux and continuum errors have been propogated. Thus the two 
absorption lines are significant at $3.5\sigma$ and $3.8\sigma$, with 
corresponding null probabilities of $4.6\times10^{-4}$ and 
$1.4\times10^{-4}$. However these are single trial probabilities for 
a line at an a-priori known energy. Taking into account the number of bins $N$ 
searched for line features in the 3-7 keV range, then the null detection 
probability $P$ will be $P=1-(1-P_{1})^{N}$, where $P_{1}$ is the 
single trial probability from above. As there are 48 resolution bins 
in the 3-7 keV range, the null probabilities for the 4.2 and 4.9\,keV lines 
are 0.022 and 0.0067 respectively, in excellent agreement with the 
Monte-Carlo estimates. 

\section{Discussion}


Two absorption lines are observed at 4.22 and 4.93\,keV in the \PG\ LETG 
spectrum, with a low probability that both are false detections.  
Although the lines are likely to be associated with redshifted 
Fe K$\alpha$, we first consider whether the lines 
could originate from other elements at lower velocities.  
The absorption line rest 
frame energies are $4.56\pm0.03$~keV and $5.33\pm0.03$~keV. 
The closest transitions 
to those energies are Sc~\textsc{xxi} K$\alpha$ (at 4.53\,keV) or 
V~\textsc{xxiii} (5.43\,keV) respectively. The 
abundances of Sc or V are extremely low, $\sim1000\times$ less than Fe.  
Even if the abundances of were enhanced through 
spallation of iron nuclei by energetic protons 
on the disk surface \citep{Skibo97}, the strengths of 
these lines would still be considerably weaker 
($<<10\%$) than due to Fe K$\alpha$. 
One other possibility is that the absorption originates 
from lower Z elements such as Si, S or Ca in a 
relativistic outflow. Blue-shifted lines with 
$v\sim0.1c$ have been detected in \PG\ in a 
previous XMM-Newton observation by \citet{Pounds03}. In this case 
Ca~\textsc{xx} at 4.10 keV would appear the most reasonable 
identification for the lower energy line, requiring an 
outflow velocity of 32000\,km\,s$^{-1}$. However the  
Ca~\textsc{xx} line is still likely to be weak, as the abundance ratio is 
${\rm Fe/Ca}\sim20$ \citep{Anders89}. Si or S are more abundant than Ca, 
however if the lines originate from Si~\textsc{xiv} K$\alpha$ and 
S~\textsc{xvi} K$\alpha$, then the  
outflow velocities are 0.68c and 0.62c respectively. The 
kinetic power of such a fast flow would be a factor of $\sim40$ greater 
than calculated by \citet{Pounds03}, probably exceeding 
$10^{46}$~erg~s$^{-1}$ and the bolometric output of the quasar. 
Such a highly energetic flow seems implausible, unless the opening 
angle of the outflow is small. 

The most likely scenario is that the lines originate from 
highly redshifted iron K$\alpha$. 
Indeed there have been a small number of previous 
claims of redshifted Fe K absorption lines in other AGN. \citet{Nandra99} 
first claimed the possibility of redshifted iron K-shell 
absorption, through \asca\ observations of the Seyfert 1 NGC 3516.  
However recent Chandra and XMM-Newton observations \citep{Turner02} 
have showed that this is probably due to a strong narrow K$\alpha$ 
core at 6.4\,keV and variable emission lines at $<6.4$\,keV. 
Other claimed examples include 
a 6.2 keV line in the quasar 
1E\,1821+643 with Chandra HETG \citep{Yaqoob05} and a 
weak redshifted line in the quasar Q 0056-363 \citep{Matt05}. 
Most recently the detection of transient 
redshifted Fe absorption lines in Mrk 509 was claimed in 
Beppo-SAX data, with a velocity shift of $v=0.21c$ 
\citep{Dadina05}. 

If the lines in \PG\ correspond to H-like iron 
(i.e. Fe~\textsc{xxvi} K$\alpha$ at 6.97 keV) then the 
redshifted velocities are 
0.26c and 0.40c, while if the lines correspond to He-like iron 
(Fe~\textsc{xxv} K$\alpha$ resonance at 6.70\,keV), the velocities are 
0.22c and 0.37c. Note that the separation of the two lines requires two 
distinct velocity components, as the higher order Fe K$\beta$ lines are 
undetectable in this spectrum. It is unlikely that the lines 
result from a lower ionization than Fe~\textsc{xxv}, as then a series 
of strong L-shell Fe lines and edges would be detected at 
lower energies, which is not the case. Furthermore the absorption lines  
cannot be from near neutral Fe (Fe~\textsc{i-xvii}) at 6.4 keV, because no 
strong K$\alpha$ absorption line would be observed, as the L-shell would be 
fully populated. 
 
The two possible causes for the relativistic shifts are 
either gravitational redshift of photons 
near the massive black hole or infall of matter onto the black hole. In the 
former scenario, the absorbing matter would have to be located at 
$3.5\pm0.5R_{\rm g}$ and $4.8\pm0.5R_{\rm g}$ respectively 
from a Schwarzschild black hole (where $R_{\rm g}=GM/c^{2}$ 
is one gravitational radius). This is within 
the last stable orbit around a Schwarzschild black hole and 
requires a Kerr metric \citep{Thorne74} for the matter to exist in 
a stable orbit within $6R_{\rm g}$. For a maximal pole-on Kerr black hole, the 
radii derived are then $3.2R_{g}$ and $4.6R_{g}$ 
respectively. Alternatively, the matter may be infalling onto the 
black hole, in which case a maximum velocity of $\sim0.38c$ (in 
the frame of the observer) can be achieved at $6R_{\rm g}$ for a Schwarzschild 
metric.

Modeling the two absorption line
systems with Xstar \citep{Kallman96}, requires a large column density 
of $N_{\rm H}=4.0^{+3.7}_{-1.9}\times10^{23}$~cm$^{-2}$ (assuming a 
turbulence of $\sigma_{\rm v}=1000$\,km\,s$^{-1}$ and solar abundances).  
The ionization parameter is log\,$\xi=3.9\pm0.3$~erg~cm~s$^{-1}$, 
with most of the absorption arising from Fe~\textsc{xxvi} K$\alpha$. 
The equivalent column density of Fe~\textsc{xxvi} is 
$\sim7\times10^{18}$\,cm$^{-2}$, while the optical depth at the 
K-shell edge is $\tau<0.1$, i.e. the matter is optically-thin to the 
continuum. The velocity of the two absorption 
systems are $v=(0.28\pm0.01)c$ and $v=(0.42\pm0.01)c$, while the 
line of sight covering fraction is $>50$\%.  
From the above parameters we can derive the physical 
properties of the gas. A black hole mass of $4\times10^{7}$~\Msun\ 
is assumed \citep{Kaspi00}, an ionizing luminosity  
of $L\sim10^{44}$~erg~s$^{-1}$ and a radial distance of 
$R\sim10^{14}$\,cm (i.e. 
a few $R_{\rm g}$).  
The ionization parameter is defined as $\xi=L/nR^{2}$ (n is the 
electron density), while one can define a filling factor $f$ as 
$f=\Delta R/R=N_{\rm H}/nR$. A 
density of $n\sim10^{12}$\,cm$^{-3}$ is calculated, 
with a filling factor of $\sim$1\% and a cloud size of 
$\sim10^{12}$\,cm. Note that these should be considered
order of magnitude estimates, due to the uncertainties in $R$ and 
black hole mass. The matter may be clumped, perhaps  
from density perturbations on the disk or 
from localized filaments within the infalling matter. 

A fast outflow of $\sim0.1c$ has already been detected in an earlier 
XMM-Newton observation of \PG\ \citep{Pounds03} which is confirmed from the 
analysis of the soft X-ray LETG data (Reeves et al. 2005, in preparation). 
One possibility is that 
part of this outflow does not escape the gravitational potential 
of the black hole \citep{Murray98, Proga00}. 
If material is launched with $v=0.1c$ in \PG, then 
matter within a radius of $R<100R_{\rm g}$ may not escape the system. 
Thus it is possible to produce 
both red and blue-shifted lines in \PG, indeed in Mrk 509 both 
inflow and outflow also appear to be present \citep{Dadina05}. 
From above, the mass infall rate required is 
$\dot{M} \sim \pi f nR^{2}vm_{\rm p} \sim \pi Lvf m_{\rm p}/\xi 
\sim10^{25}$\,g\,s$^{-1}$ 
or $\sim0.1$\,\Msun\,yr$^{-1}$. The kinetic power is then a 
few~$\times10^{44}$~erg~s$^{-1}$ for an infall velocity of $v=c/3$. 
Interestingly this is similar to the X-ray luminosity of \PG. 
Indeed a similar ``aborted jet'' model 
has been proposed by \citet{Ghisellini04}, where matter in an outflow 
or jet fails to escape the gravitational potential well. 
The hard X-ray emission can be produced by shocks and collisions 
within the flow. 

Alternatively the absorption may originate from matter 
near the highly ionized inner disk
surface, with the redshift being largely gravitational in origin. 
\citet{RF00} proposed such a model to explain the 
absorption line in NGC 3516, although the lines will be broad 
unless the matter is clumped.   
Monitoring the line
variability will help to reveal its origin, for instance 
it may possible to track the infall of a 
dense clump of matter towards the black hole. This bodes 
well for future longer observations of \PG, with XMM-Newton or Chandra,  
to track the line variability. Regardless of the mechanism, 
the large redshift and high covering fraction 
imply that most of the X-ray emission 
in \PG\ originates from a compact region within a few $R_{\rm g}$ of
the black hole.

\clearpage

\begin{figure}
\begin{center}
\rotatebox{-90}{
\epsscale{0.7}
\plotone{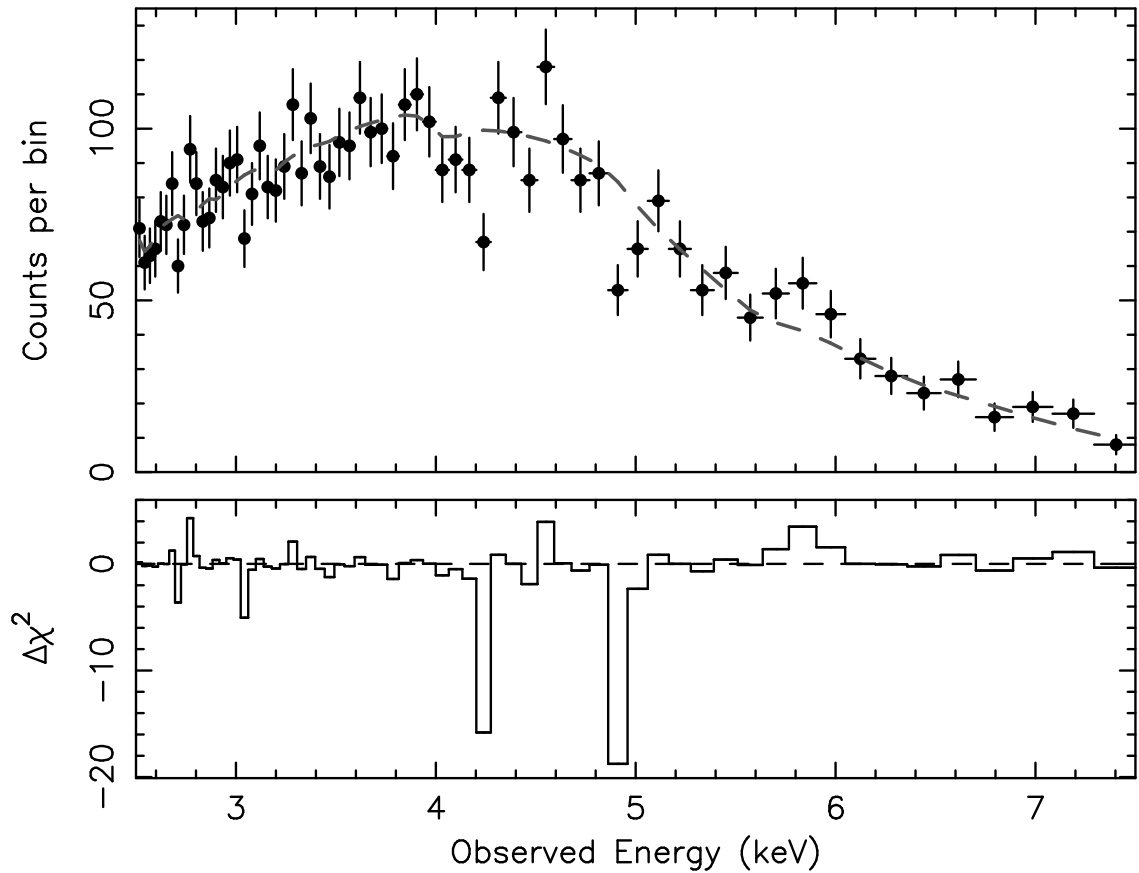}}
\caption{The upper panel shows the first order Chandra LETG spectrum of 
PG 1211+143 above 2 keV; data points are shown as filled 
circles and $1\sigma$ error bars as vertical lines. The best fitting 
continuum model, folded through the LETG 
response, is shown as a dashed line. 
The spectrum has been binned to the FWHM LETG spectral resolution 
and is plotted in the observed frame. The lower panel plots the 
$\chi^{2}$ residuals between the data and the continuum model, negative 
residuals correspond to a deficit of counts. Two 
absorption features are apparent at 4.2 and 4.9 keV respectively.}
\end{center}
\end{figure}

\clearpage

\begin{figure}
\begin{center}
\rotatebox{-90}{
\epsscale{0.7}
\plotone{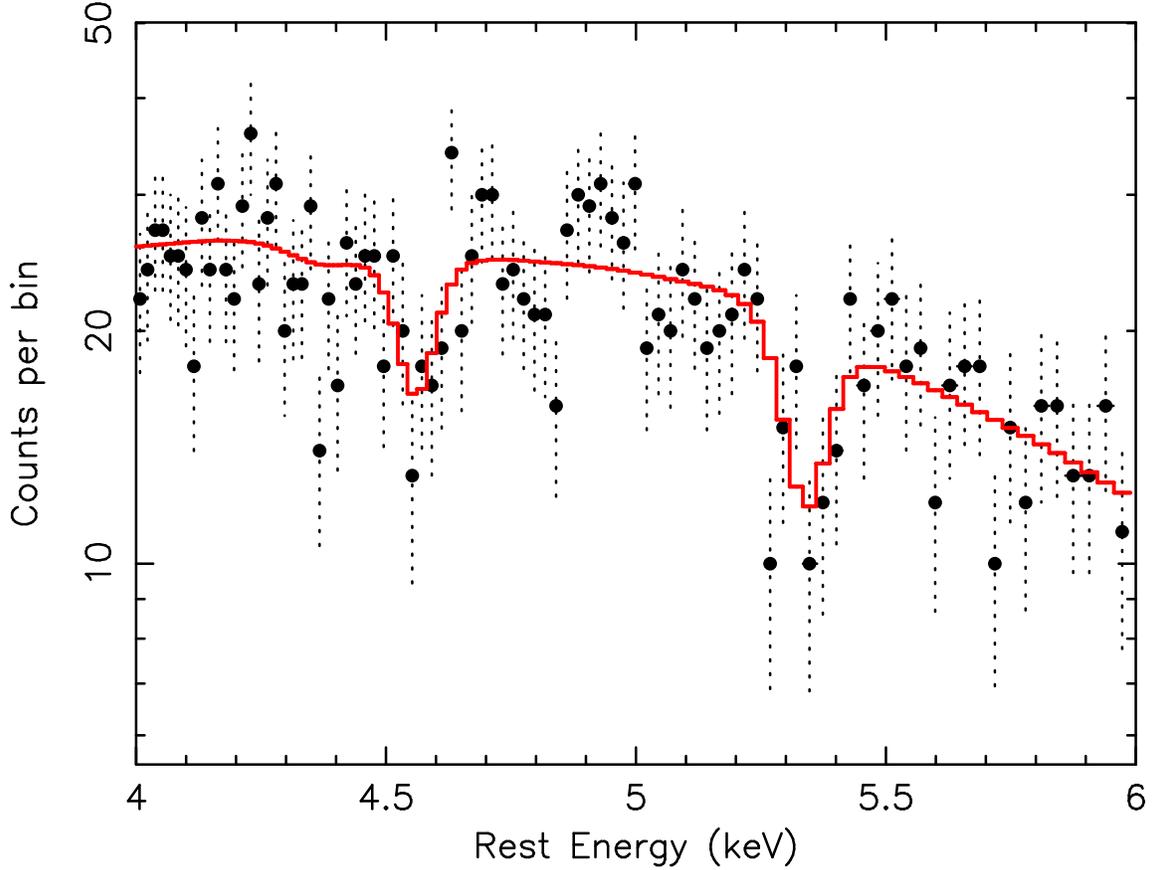}}
\caption{LETG spectrum of PG 1211+143 between 4-6 keV, binned at 
the full LETG resolution of $\Delta\lambda=0.0125$~\AA\ 
(equivalent to 1/4 FWHM resolution per bin). Data are shown as filled circles, 
with dashed lines showing the $1\sigma$ error bars and  
the best fit absorption line model as a solid line. The spectrum has been 
transformed into the rest frame of PG 1211+143. If identified with  
Fe XXVI K$\alpha$, the redshift of each line would 
correspond to velocities of 0.26c and 0.40c. 
Note that the relative separation of the lines 
requires that two separate velocity components are present.}
\end{center}
\end{figure}


\clearpage

\begin{figure}
\begin{center}
\rotatebox{-90}{
\epsscale{0.7}
\plotone{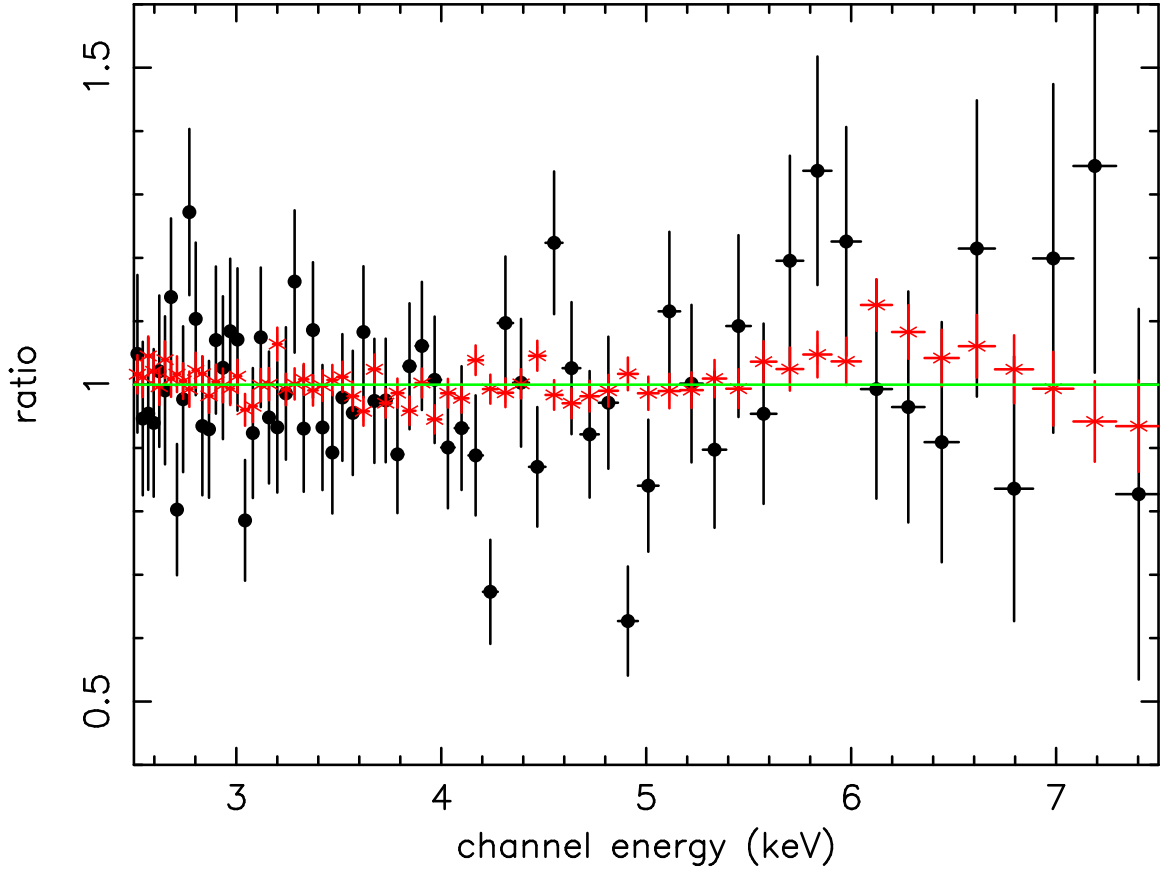}}
\caption{Data/Model ratio residuals of the PG 1211+143 (circles) 
and 3C 273 (stars) LETG spectra to power-law continuum models fitted 
above 2 keV. The data were binned at the FWHM resolution of the LETG. 
The absorption lines apparent between 4-5 keV in PG 1211+143 do not 
appear in the 3C 273 spectrum, ruling out calibration effects.}
\end{center}
\end{figure}

\clearpage

\begin{deluxetable}{lcccc}\label{tab:line}
\tabletypesize{\normalsize}
\tablecaption{Table of iron line spectral fits.}
\tablewidth{0pt}
\tablehead{
\colhead{Parameter} & \colhead{Mean} & \colhead{Obs1} &   
\colhead{Obs2} & \colhead{Obs3}} 

\startdata
Date\tablenotemark{a} & 21/06/2004 & 21/06/2004 & 
23/06/2004 & 25/06/2004 \\

Time\tablenotemark{a} & 03:09:04 & 03:09:04 & 03:57:43 &18:30:27 \\  

Exposure\tablenotemark{b} & 133.6 & 42.7 & 48.2 & 42.7 \\

$\Gamma$\tablenotemark{c} & $1.99\pm0.06$ & $1.87\pm0.14$ & $2.00\pm0.10$ & 
$1.93\pm0.10$ \\

F$_{2-10}$\tablenotemark{d} & 6.1 & 4.0 & 6.6 & 7.3 \\

Energy\tablenotemark{e} & $4.22\pm0.03$ & 4.22\tablenotemark{f} & 
$4.22\pm0.03$ & 4.22\tablenotemark{f} \\

EW\tablenotemark{g} & $38\pm16$ & $<25$ & $54\pm24$ & $<46$ \\

Energy\tablenotemark{e} & $4.93\pm0.03$ & 4.93\tablenotemark{f} & 
$4.89\pm0.04$ & 4.93\tablenotemark{f} \\

EW\tablenotemark{g} & $62\pm23$ & $<21$ & $69\pm37$ & $<57$ \\



\enddata


\tablenotetext{a}{Start date and time (in UT) for observation.}
\tablenotetext{b}{Exposure time in ks.}
\tablenotetext{c}{Photon index above 2 keV.}
\tablenotetext{d}{2-10 keV flux in units $10^{-12}$~ergs~cm$^{-2}$~s$^{-1}$.}
\tablenotetext{e}{Observed line energy in keV.}
\tablenotetext{f}{Denotes model parameter is fixed in fit.}
\tablenotetext{g}{Absorption line equivalent width in eV.}

\end{deluxetable}
\end{document}